\documentclass[namedreferences]{solarphysics}
\usepackage[hyperref,optionalrh]{spr-sola-addons} % For Solar Physics 
\usepackage[optionalrh]{spr-sola-addons} % For Solar Physics
\usepackage{graphicx,color}        % For eps figures, newer & more powerfull
\usepackage{amssymb}         % useful mathematical symbols
\usepackage{color}           % For color text: \color command
\usepackage{url}             % For breaking URLs easily trough lines
\usepackage{epstopdf}

\newcommand{\kms}{km\,s$^{-1}$}

\begin{document}

\begin{article}

\begin{opening}

\title{Kinematics and Energetics of the EUV Waves on 11 April 2013}

\author[addressref={aff1},corref,email={aarti.solarphysics@gmail.com}]{\inits{Aarti}\fnm{Aarti}~\lnm{Fulara}\orcid{0000-0003-2973-5847}}
\author[addressref=aff1]{\inits{Ramesh}\fnm{Ramesh}~\lnm{Chandra}\orcid{0000-0002-3518-5856}}
\author[addressref={aff2,aff3}]{\inits{P.F.}\fnm{P.F.}~\lnm{Chen}\orcid{0000-0002-7289-642X}}
\author[addressref=aff4]{\inits{Ivan}\fnm{Ivan}~\lnm{Zhelyazkov}\orcid{0000-0001-6320-7517}}
\author[addressref=aff5]{\inits{A.K.}\fnm{A.K.}~\lnm{Srivastava}\orcid{0000-0001-5152-8719}}
\author[addressref=aff6]{\inits{Wahab}\fnm{Wahab}~\lnm{Uddin}\orcid{0000-0003-2829-5907}}

\address[id=aff1]{Department of Physics, DSB Campus, Kumaun University, Nainital 263001, India}
\address[id=aff2]{School of Astronomy \& Space Science, Nanjing University, 163 Xianlin Ave, Nanjing 210023, China}
\address[id=aff3]{Key Laboratory of Modern Astronomy \& Astrophysics (Nanjing University), Ministry of Education, Nanjing 210023, China}
\address[id=aff4]{Faculty of Physics, Sofia University, 1164 Sofia, Bulgaria}
\address[id=aff5]{Department of Physics, Indian Institute of Technology (BHU), Varanasi 221005, India}
\address[id=aff6]{Aryabhatta Research Institute of Observational Sciences, Manora Peak, Nainital 263002, India}

\runningauthor{A.\ Fulara {\it et al.}}
\runningtitle{EUV Wave on 11 April 2013}

\begin{abstract}
: In this study, we present the observations of extreme-ultraviolet (EUV) waves 
associated with an M6.5 flare on 2013 April 11. The event was observed by 
\emph{Solar Dynamics Observatory\/} (SDO) in different EUV channels. The flare 
was also associated with a halo CME and type II radio bursts. We observed both 
fast and slow components of the EUV wave. The speed of the fast 
component, which is identified as a fast-mode MHD wave, varies in the range 
from 600 to 
640~\kms, whereas the speed of the slow-component is ${\approx}$140~\kms. We 
observed an unusual phenomenon that, as the fast-component EUV wave passes 
through two successive magnetic quasi-separatrix layers (QSLs), two stationary 
wave fronts are formed locally. We propose that part of the outward-propagating 
fast-mode EUV wave is converted into slow-mode magnetohydrodynamic waves, which 
are trapped in local magnetic field structures, forming successive 
stationary fronts. Along the other direction, the fast-component EUV wave also 
creates oscillations in a coronal loop lying ${\approx}$225 Mm away 
from the flare site. We have computed the energy of the EUV wave to be of the order of $10^{20}$~J.

\end{abstract}

\end{opening}
%%%%%%%%%%%%%%%%%%%%%%%%%%%%%%%%%%%%%%%%%%%%%%%%%%%%%%%%%%%%%%%%%%%%%%%%%

\section{Introduction}
     \label{S-Intro}
Globally propagating disturbances traveling through the corona first came into existence by the observation made by the \emph{Extreme-Ultraviolet Imaging Telescope\/} (EIT) 
(EIT: \opencite{Del95}) on board the \emph{Solar and Heliospheric Observatory\/}  (SOHO) (SOHO: \opencite{Domingo95}), thus known as ``EIT waves.''  Being clearly visible in various EUV wavelengths, they are also termed as EUV waves.  EUV waves are characterized by bright and diffuse fronts  which can sometimes travel the whole solar disk (\opencite{Moses97}; \opencite{Thompson98}). Their speeds range from a few tens \kms up to more than 1000~\kms{} (\opencite{Thompson09}; \opencite{Nitta13}; \opencite{Muhr14}; \opencite{Long17}; \opencite{Zheng18}).

 Coronal mass ejections play an important role in the generation of EUV waves. Since CMEs are best detected above the solar limb and EUV waves are best seen on the disk, it is always difficult to clarify their association. A statistical analysis by \inlinecite{Bie02} revealed that all the clearly-identified EUV waves are associated with CMEs, though only 58\% of all EUV waves, including faint events, are accompanied by CMEs. In contrast, only 20\% of the CMEs have waves associated with them (\opencite{Thompson09}). \inlinecite{Muhr14} did a more detailed study on the relationship between EUV waves and CMEs and found that 95\% of 60 EUV waves in their sample are associated with CMEs. Nowadays, it is widely believed that EUV waves are intimately associated with CMEs, rather than solar flares. In particular, it was proposed that the CME frontal loop is cospatial with the slow-component of EUV waves, whereas the fast-component EUV wave corresponds to the piston-driven shock wave straddling over the CME (\opencite{Chen09}). The cospatiality between the CME flank and the slow-component EUV wave was later confirmed by several authors (\opencite{Chen09}; \opencite{Attrill09}; \opencite{dai10}; \opencite{Zhou17}). It is noted, however, that since EUV waves and CMEs are often observed in different fields of view, their association still requires more detailed studies.
	
Regarding the nature of EUV waves, several models have been proposed. Initially 
it was widely believed that EUV waves are fast-mode magnetohydrodynamic 
(MHD) waves (\opencite{Thompson98}; \opencite{Wang2000}; \opencite{Wu01}; 
\opencite{Pat09}; \opencite{Schmit10}), which was seconded by the 
observational features like wave reflection, refraction, and transmission 
(\opencite{Gopal09}; \opencite{Kienreich13}; \opencite{Veronig08}; 
\opencite{Long08}). However, contrary to this, slower EUV waves, whose 
speeds are even smaller than the coronal sound speed, have also been observed. 
These waves cannot be explained by the fast-mode wave model, and have been 
proposed to be due to Joule heating (\opencite{Del07}), successive reconnection 
(\opencite{Attrill07}), or slow-mode waves (\opencite{Mei12}). In addition, the 
existence of stationary fronts (\opencite{Del07}; \opencite{Chandra09}) 
challenged the fast-mode wave model for all EUV waves. To reconcile all these 
discrepancies, \inlinecite{Chen02} and \inlinecite{Chen05} proposed a hybrid 
wave theory, \emph{i.e.} there are two types of EUV waves associated with one coronal 
mass ejection (CME) event, where the outer sharp wave front is a fast-mode MHD 
wave or shock wave and the inner diffuse front with a smaller speed is an 
apparent wave produced by successive stretching of magnetic field lines. 
Such a two-component EUV wave scenario was further supported by 
three-dimensional simulations (\opencite{Cohen09, Downs12}). With the high-cadence data, the co-existence of both the fast-mode wave and the slow-component EUV wave has been reported by many authors (\opencite{Chen11}; \opencite{Cheng12}; \opencite{Asai12}; \opencite{Kumar13}; \opencite{Chandra16}; \opencite{Zong17}; \opencite{Chen17}). Recent reviews on EUV waves can be found in \inlinecite{Warmuth07}, \inlinecite{Wills09}, \inlinecite{Warmuth10}, \inlinecite{Gallagher11}, \inlinecite{Zhukov11}, \inlinecite{Liu14}, \inlinecite{Warmuth15}, and \inlinecite{Chen16b}.

\inlinecite{Del99} first reported the existence of stationary brightening in EUV images. They proposed that the stationary fronts are due to Joule heating of the electric currents generated near magnetic quasi-separatrix layers (QSLs) as the magnetic field lines are opening during a CME, and it was used as evidence to argue against the fast-mode wave model for EUV waves. Their work invoked \inlinecite{Chen02} to propose the magnetic fieldline stretching model for EIT waves, and this model can naturally explain why the slow-component EUV wave, \emph{i.e.} the non-wave component stops at a magnetic QSL (\opencite{Chen05}). On the other hand, \inlinecite{Chandra16} for the first time reported a different scenario in observations, \emph{i.e.} in addition to a stationary front being the decelerating slow-component EUV wave, another stationary front is generated as the fast-component EUV wave passes through a magnetic QSL. The fast-component EUV wave continues its journey but with much reduced intensity. It was proposed by \inlinecite{Chen16} \emph{via} MHD simulations that the new stationary front is a new-born slow-mode MHD wave, which is converted from the incident fast-mode EUV wave. Very recently, \inlinecite{Zong17} and \inlinecite{Chandra18} found in observations that fast-mode EUV waves can indeed be converted into slow-mode waves when passing through helmet streamers, whose boundary corresponds to a magnetic QSL. More interestingly, \inlinecite{Del07} revealed multiple stationary fronts, which were visible in both EUV and H$\alpha$ wavelengths. However, owing to the low cadence of the EIT telescope, they cannot tell whether the stationary fronts are produced by the opening of magnetic field lines or by the perturbation of the ambient coronal magnetic field while a magnetosonic wave passes through them. 

The fast-component EUV waves are considered to be coronal shock waves. Shock waves signify the appearance of type II radio bursts. Type II radio bursts appear as strips of enhanced radio emission slowly drifting from high to low frequencies in the radio dynamic spectra. \inlinecite{Bie02} pointed out that a type II radio burst is a sufficient but not a necessary condition for an EUV wave. Recently, \inlinecite{Long17} did a statistical analysis of the events with EUV waves and type II radio bursts and found that only 40\% of the wave events have type II bursts associated with them. Also, they reported that there is no clear relation between the velocity of the wave and the drift speed of the type II radio burst. Thus, this lack of clear relationship may be again attributed to the fact that EUV waves propagate in the low corona while radio bursts are related to the upper corona (\opencite{Mann03}; \opencite{Pat09}). Hence, their correlation is still controversial and needs more detailed studies. 

Regarding the energetics of EUV waves, there are still controversies.  Their energy varies from $10^{16}$ to $10^{24}$~J. Therefore, studying the energetics of EUV waves, still awaits more efforts. A few papers addressing the energetics of EUV waves are summarised as follows: \inlinecite{Ballai05} computed the energy of an EUV wave based on the parameters of loop oscillations. The loop oscillations were produced by the EUV wave. Their calculations were based on typical coronal parameters such as the phase speed of the wave generated in the loop, sound speed, cusp speed (which is slow-mode speed in the external region) and density ratio inside and outside the loop.  Their assumption was based on the fact that all the energy of the wave is transferred to the oscillating loop and this computed energy is the lower limit for the EUV wave. Based on the method proposed by \inlinecite{Ballai05}, \inlinecite{Ballai07} did a statistical analysis of 14 EUV wave events and obtained the energy of EUV waves in the range $10^{16}$--$10^{19}$~J . \inlinecite{Gilbert08} studied a wave-filament interaction and computed the maximum total kinetic energy involved in the interaction. Their resulting energy is in the range of ${\approx}10^{19}$--$10^{20}$~J.  \inlinecite{Pat12} adopted a different approach to compute the total energy of EUV waves. They suggested that the energy of the EUV wave can be considered as the sum of three terms, \emph{i.e.} the kinetic energy flux, the radiative loss flux, and the coronal thermal conduction flux. Their computed energy is of the order of $10^{22}$~J. More recently, \inlinecite{Long15} also estimated the energy of the coronal waves using an approximation for shock waves which propagate in a region of variable density, and their energy of the EUV wave turns out to be ${\approx}10^{24}$~J.

In this paper we present the observations of an EUV wave event on 2013 April 11 originating from the active region NOAA AR 11719. Section \ref{obs} illustrates the observational data sets and the general overview of the event. In Section \ref{res}, we describe the kinematics and energetics of the EUV wave. The associated CME and type II radio bursts are described in Section \ref{cmes}. Finally in Section \ref{sum}, we discuss and conclude our results.

%%%%%%%%%%%%%%%%%%%%%%%%%%%%%%%%%%%%%%%%%%%%%%%%%%%%%%%

\section{Observational Data Sets and General Overview of the Event} \label{obs}

The flare and the associated EUV wave on 2013  April 11 are well observed by the \emph{Atmospheric Imaging Assembly\/} (AIA) (AIA: \opencite{Lemen12}) on board the \emph{Solar Dynamics Observatory\/} (SDO) (SDO: \opencite{Pesnell12}) satellite, which observes the full Sun at different wavelengths in EUV and UV with high spatial (0.6$''$) and temporal (12~s) resolutions.  For the present study, we use AIA 171 \AA, 193 \AA, and 1600~\AA\ data. To see the chromospheric signatures of the flare and EUV waves, we use H$\alpha$ data from the \emph{Global Oscillation Network Group\/} (GONG) instruments.
GONG observes the full Sun in H$\alpha$ with a cadence of 1 min and a spatial resolution of 2$''$. For the associated type II radio bursts and the associated CME, we use the data from the \emph{Hiraiso Radio Spectrograph\/} (HiRAS) (HiRAS: \opencite{Kondo95}) and the \emph{Large Angle and Spectrometric Coronagraph} (LASCO) (LASCO: \opencite{Brueckner95}) on board the SOHO satellite.

The active region NOAA AR 11719 appears near the east limb on 2013 April 7 and turns behind the west limb on 2013 April 17. The flare starts at $\approx$06:55~UT on April 11 and is classified as M6.5 class according to the GOES X-ray flux. Before the flare onset, a sigmoidal structure is visible in AIA 94~\AA{} wavelength. The detailed study of this sigmoidal formation was done by \inlinecite{Vema15} and \inlinecite{Joshi17}. Figure \ref{fla} presents a multiwavelength view of the flare in different AIA channels and in GONG H$\alpha$. The flare begins as two bright kernels, which gradually expand to form two reverse J-shaped ribbons. Figure \ref{fla}c shows the peak phase of the flare at 07:10~UT in H$\alpha$. The flare ribbons show separation from each other, which is typical for two-ribbon flares. The reverse J-shaped ribbon in the west indicates negative helicity in the active region as reported in several observations (\opencite{Chandra11}; \opencite{Sch15}; \opencite{Janvier17}). In Figure \ref{fla}d, we overplot the HMI magnetic field contours on the AIA 1600~\AA\ image.  The structure of the flare ribbons as well as the existence of the sigmoid, both are consistent with negative helicity in the active region. The active region is located in the northern hemisphere which is dominated by negative helicity. Therefore, this active region follows the hemispheric rule (\opencite{Ouyang17}).

\section{Results} \label{res}

\subsection{Kinematics of the EUV Waves}
Figure \ref{wave} presents the propagation of the wave in the running difference images in AIA 193~\AA.  The difference images are created by subtracting the previous image as indicated in each image. The first trace of the wave is found at $\approx$07:04~UT, and the EUV wave is quasi-circular in shape. It emanates from the active region and propagates mainly in the southeast direction away from the active region. As the wave progresses, its front becomes more and more diffuse. By 07:11 UT the wave almost reaches the east limb. As seen from Figure \ref{wave}, the EUV wave is accompanied by dimmings in the wake. The dimmings are formed in the region enclosed between the flare site and the boundary of the EUV wave. Dimmings are often interpreted as regions of evacuation of coronal mass during a CME (\opencite{Sterling97}; \opencite{Wang02}; \opencite{Harra03}; \opencite{Zhukov04}; \opencite{Jin09}). However, there is distinction among the stronger core dimmings which are attributed to the footpoints of the flux rope (\emph{e.g.} \opencite{Sterling97}; \opencite{Webb2000}), the fainter expanding dimmings, which are observed to trail behind the slow-component EUV wavefront (\emph{e.g.} \opencite{Del99}; \opencite{Wills99}), and the rarefaction in the wake of the fast-component EUV wave (\emph{e.g.} \opencite{Muhr11}, \opencite{Lulic13}). What we see in Figure \ref{wave} are mainly the expanding dimmings and the rarefaction as indicated by the time-distance diagrams later shown in Figures \ref{qsl} and \ref{twowave}, whereas the core dimmings are localized at the boundaries of the source active region.
 
In order to study the kinematics of the observed EUV waves, after examining their propagation along several artificial slices in different directions from the flare site, we select two representative slices, where the propagation of EUV waves is clearly visible. We label them Slice 1 and Slice 2, respectively. Let us first discuss about Slice 1. This slice extends toward the east direction (marked by the black curved line in Figure \ref{qsl}a). The corresponding time--distance diagram of the base-difference AIA 193 \AA\ intensity is shown in Figure 3b. A prominent feature in Figure \ref{qsl} is that a fast-moving wave propagates toward the east direction. The calculated speed is $\approx$640 km s$^{-1}$, which is several times larger than the typical sound speed in the corona and is typical for the fast-component EUV waves (\opencite{Chen16b}). That is the reason why we call it a fast-mode EUV wave. Although the slow-component EUV wave is not as discernable here as in \inlinecite{Chen11} and \inlinecite{Kumar13}, one can still see two patchy brightenings as demonstrated by \inlinecite{Guo15b}. The transiting of the sequential brightenings forms a wavelike pattern, which can also be explained by the magnetic fieldline stretching model. Moreover, it is seen that these sequential patchy brightenings bound the expanding dimmings, as expected from the magnetic fieldline stretching model. Another interesting feature in Figure \ref{qsl} is that when the fast-mode EUV wave propagates outward, two stationary brightenings are generated, which are marked as SB$_1$ and SB$_2$, respectively. The lifetimes of these stationary brightenings are tens of minutes. 

In Figure \ref{pfss} we compare the locations of the two stationary brightenings with the coronal magnetic field, which is extrapolated from the photospheric magnetogram with the potential field source surface (PFSS) model. It reveals that the two stationary brightenings along Slice 1 are both located at magnetic QSLs, where magnetic connectivity changes abruptly.

We also trace the propagation of the EUV waves along Slice 2, which is along the southeast direction as indicated in Figure \ref{twowave}a. The corresponding time--distance diagram of the base-difference AIA 193 \AA\ intensity is presented in Figure 5b. Along Slice 2, we find the co-existence of two EUV waves, \emph{i.e.} a fast component and a slow component. According to the magnetic fieldline stretching model, the fast-component EUV wave is a fast-mode MHD wave or shock wave, whereas the slow-component EUV wave is an apparent motion generated by the successive stretching of magnetic field lines pushed by an erupting flux rope (\opencite{Chen02}). The speed of the fast-mode wave is $\approx$600 km s$^{-1}$, which is again several times higher than the coronal sound speed and this speed is similar to the speed of fast-mode wave along Slice 1. The speed of the slow-component EUV wave is $\approx$140 km s$^{-1}$, which is about 4.3 times smaller than that of the fast-component EUV wave.

\subsection{Energetics of the EUV Wave}

Figure \ref{loop} shows an AIA 171~\AA{} image at 07:20~UT. We can see a coronal loop L$_1$ indicated by the white arrow. As the EUV wave propagates along Slice 2, it encounters the loop system L$_1$. The L$_1$ loop starts to 
oscillate along the Slice 2 direction. We observe strong and clear oscillations of this loop. This loop system is ${\approx}$225~Mm away from the flare site. \inlinecite{Guo14} also studied the oscillations of this particular loop. For the computation of the electron density inside the coronal loop, they calculated the background subtracted EUV fluxes in the six AIA wavelengths. Using these they derived the average electron density inside the loop to be $5.1\pm0.8\times 10^{8}$~cm$^{-3}$ and the temperature to be $0.65\pm 0.06$~MK. \emph{Via} coronal seismology they derived the magnetic field strength, which is $B_\mathrm{i} = 8.2$~G. We created a time-distance plot using a series of 171~\AA\ images. Figure \ref{slice}a shows the position of the slice along which the 171 \AA\ intensity distribution is extracted. Figure \ref{slice}b shows the time-slice diagram illustrating the loop oscillations. As seen from the figure, the first shift of the loop is in a direction away from the active region. Initially, the amplitude of the oscillations is large, but then decays. Using this time-distance diagram, we create an intensity-time plot for the oscillations which is shown in Figure \ref{wavelet}. From the plot, we can see that the the maximum deflection of the loop is $1.9\times10^{4}$~km. We compute the period of the loop oscillations using the wavelet analysis. Wavelet analysis helps us study the time-dependent period in the observed light-curves \cite{Torrence98}. The period of the loop oscillations revealed from the wavelet analysis is 541~s. The power spectrum of the wavelet analysis is shown in Figure \ref{wavelet}. Along with other computed observational parameters such as the length of the loop, magnetic field, temperature, and the radius of the loop, we also calculate the maximum and intermediate deflections of the loop and note down the respective times at which these deflections take place.

\begin{table}[t]

\medskip
\centering
\begin{tabular}{l l}
\hline
Input parameters & Value\\
\hline
Length of the loop (\emph{L}) & 198~Mm\\
Coronal magnetic field (\emph{B}) & 8.23~G (\opencite{Guo14})\\
Number density (\textit{$n_i$}) & $5.1\times10^{8}$~cm$^{-3}$ (\opencite{Guo14})\\
Radius of the loop (\emph{R}) & 4.3~Mm\\
Temperature (\emph{T}) & 0.65~MK (\opencite{Guo14})\\
Mass density inside the loop ($\rho_\mathrm{i}$) & $8.16\times10^{-13}$~kg\,m$^{-3}$ \\
Mass density outside the loop ($\rho_\mathrm{e}$) & $6.27\times10^{-13}$~kg\,m$^{-3}$\\
Maximum deflection of the loop ($x_\mathrm{max}$) & 1.9~Mm\\
Intermediate deflection of the loop (\textit{$x_1$}) & 1.2~Mm\\
Time at which maximum deflection occurs ($t_\mathrm{max}$) & 540~s \\
Time at which intermediate deflection occurs (\textit{$t_1$}) & 1080~s \\
Sound speed outside the loop ($c_\mathrm{se}$) & $94.6$~\kms\\
Alfv\'en speed inside the loop ($v_\mathrm{Ai}$) & $791.6$~\kms\\
Alfv\'en speed outside the loop ($v_\mathrm{Ae}$) & $904.3$~\kms\\
Phase speed in the loop ($v_\mathrm{ph}$) & $841.6$~\kms\\
Cusp speed outside the loop ($c_\mathrm{Te}$) & $94.1$~\kms\\
Wave number (\emph{k}) & $1.5\times10^{-8}$~m$^{-1}$ \\
\hline

\end{tabular}
\caption{Observational parameters derived from the AIA observations.}
\end{table}

As mentioned in Section \ref{S-Intro}, the studies on the energy computation of the EUV waves have been done by many authors. But the difference between ours and the previous studies is that, in earlier studies (\opencite{Ballai05}; \opencite{Ballai07}; \inlinecite{Pat12}) typical coronal values of the sound speed, cusp speed, phase speed, Alfv\'en speed, temperature, and coronal density were used. To enhance our understanding on the energetics of the EUV waves, we derive these values directly from the observations. Note that some parameters are adopted from \inlinecite{Guo14}, which are also based on observations. Thus, our study gives a more realistic energy value. The observational parameters used for the energy calculation are listed in Table 1.

We calculate the energy of the EUV wave around the coronal loop L$_1$ using the following expression given by \inlinecite{Ballai05},
\begin{equation}
\label{eq:energy}
E = \frac{\pi L(\rho_\mathrm{i}R^2+\rho_\mathrm{e}/\lambda_\mathrm{e}^2)}{2}\biggl(\frac{x_\mathrm{max}-x_1}{t_\mathrm{max}-t_1}\biggr)^2,
\end{equation}
\begin{equation}
\label{eq:lambda}
\lambda_\mathrm{e}^2 = \frac{(c_\mathrm{se}^2 - v_\mathrm{ph}^2)(v_\mathrm{Ae}^2 - v_\mathrm{ph}^2)}{(c_\mathrm{se}^2 + v_\mathrm{Ae}^2)(c_\mathrm{Te}^2 -v_\mathrm{ph}^2)}k^2,
\end{equation}
where $\lambda_\mathrm{e}^{-1}$ is the decay length of perturbations outside the loop.

The basic parameters (sound and Alfv\'en speeds) used in Equation 2 are calculated from the pressure balance equation with the following input data investigated by \inlinecite{Guo14} and by us: $\rho_\mathrm{i} = 8.53 \times 10^{-13}$~kg\,m$^{-3}$, $\rho_\mathrm{e} = 6.56 \times 10^{-13}$~kg\,m$^{-3}$, $T = 0.65$~MK, and $B_\mathrm{i} = 8.2$~G, to get:
\[
	c_\mathrm{se} = 94.6~\mathrm{km}\,\mathrm{s}^{-1}
\]
along with
\[
	v_\mathrm{Ai} = 791.6~\mathrm{km}\,\mathrm{s}^{-1} \qquad \mbox{and} \qquad v_\mathrm{Ae} = 904.3~\mathrm{km}\,\mathrm{s}^{-1}.
\]
Hence, the tube/cusp speed outside the loop is
\[
	c_\mathrm{Te} = \frac{c_\mathrm{se} v_\mathrm{Ae}}{\sqrt{c_\mathrm{se}^2 + v_\mathrm{Ae}^2}} = 94.1~\mathrm{km}\,\mathrm{s}^{-1}.
\]
Since the oscillations are due to kink-mode waves, and considering that the kink mode is essentially non-dispersive, the phase velocity is equal to the kink speed. Thus, wave phase speed in Equation 2 is 
\[
	v_\mathrm{ph} = \sqrt{\frac{2}{1 + \rho_\mathrm{e}/\rho_\mathrm{i}}}v_\mathrm{Ai}.
\]
Bearing in mind that the density contrast $\rho_\mathrm{e}/\rho_\mathrm{i}$ obtained by \cite{Guo14} is equal to $0.769$, and the Alfv\'en speed inside the loop is of $791.6$~\kms, then the phase speed equals $841.6$~\kms.

With these speeds and wavenumber $k = 1.5 \times 10^{-8}$~m$^{-1}$, the value of $\lambda_\mathrm{e}^2$ calculated from Equation 1 is equal to $0.298 \times 10^{-16}$~m$^{-2}$. Taking the magnitudes of the length and radius of the loop, $L$ and $R$, as well as the maximum and intermediate deflections of the loop, $x_\mathrm{max}$ and $x_1$, along with the corresponding times $t_\mathrm{max}$ and $t_1$, we obtain, from Equation 1 the minimum energy of the EUV wave transferred to the loop is $8.88 \times 10^{18}$~J. Assuming that the wave is isotropic, we can estimate the total energy of the EUV wave by multiplying this minimum energy value, with a factor $R_\mathrm{a}$, where, $R_\mathrm{a} = 2 \pi d^{2}/A$ is the area ratio between the EUV wave dome and the coronal loop L$_1$, $d$ is the distance of the loop from the flare site, $A$ is the square of the distance between the foot points of the coronal loop (which is 145$''$). The value of $R_{a}$ is 30. Thus, the total energy of the EUV wave is $2.7 \times 10^{20}$~J.

\section{CME and Radio Observations}\label{cmes}

The M6.5 flare is associated with a full-halo CME. Figure \ref{cme} shows the C2 and C3 images observed by the LASCO coronagraph. At first, the CME appears in the C2 field of view at 07:36 UT, marked by white arrows as shown in Figure \ref{cme}. It reaches the C3 field-of-view around 09:06 UT. The speed of the CME from the linear fit is estimated to be 861.5~\kms. The acceleration of the CME from the quadratic fit is estimated to be ${-}8.07$~m\,s$^{-2}$.

Figure \ref{type} shows the radio dynamic spectrum observed by the HiRAS Radio Spectrograph on 2013 April 11. Whereas a type II radio burst is clearly discernable, a type III radio burst is fairly weak. The type III burst is observed at 06:58--06:59~UT. However, the derivative of the GOES soft X-ray light curve, representing the hard X-ray emission or the magnetic reconnection rate, peaks at 07:10 UT. It seems that the type III radio burst occurs when the derivative of the GOES soft X-ray flux starts to increase. The type II burst is observed to commence at 07:03~UT, which is 5 minutes later than the onset time of the fast-mode EUV wave. A distinct feature of the type II radio burst is that both the fundamental and harmonic components are so wide in frequency that the two components merge together, and are not separate as in most events. In order to calculate the propagation speed of the shock wave, we trace the evolution of the lower branch of the harmonic component, which is 92 MHz at 07:03 UT and 50 MHz at 07:09 UT. To derive the shock wave speed from the radio dynamic spectrum, we need to assume a density model for the corona above the source region, which is unknown. If we take the one-fold Newkirk coronal density model \cite{Newk61}, which is similar to that derived by \inlinecite{Zucc14}, the resulting shock speed would be too much smaller than the CME speed, which is probably not reasonable. According to \inlinecite{Newk61}, the two-fold model is  suitable for average active regions, we therefore take the two-fold Newkirk density model, the corresponding heights of the type II radio source region are 0.73 R$_{\odot}$ and 1.19 R$_{\odot}$ above the solar surface, and its estimated radial propagating speed is about 896~\kms.

\section{Summary and Discussion}\label{sum}
In this paper, we present the kinematics of an EUV wave event, which is associated with an \texttt{}M6.5-class flare in AR NOAA 11719 and a halo CME on 2013 April 11. Our main results are summarized as follows:

\begin{itemize}
\item Along the east direction away from the source region, 
only the fast - component EUV wave with a speed of ${\approx}$640 \kms can be seen, which corresponds to a fast-mode wave or shock wave. Along the southeast direction, two EUV waves are visible, where the fast-mode wave travels with a speed of  ${\approx}$600 \kms  and the
 slow-component EUV wave travels with a speed of ${\approx}$140 \kms. The slow-component EUV wave is 4.3 times slower than the fast-mode wave.
\end{itemize}
\begin{itemize}
\item When the fast-mode EUV wave propagates eastward, two stationary bright fronts are left behind. The locations of the two stationary fronts correspond to magnetic QSLs, where magnetic field changes its connectivity drastically.
\end{itemize}
\begin{itemize}
\item Based on the observational parameters of the oscillating loop system L$_1$, we estimated the total energy of the EUV wave, which is ${\approx}2.7\times10^{20}$~J.
\end{itemize}
\begin{itemize}
\item Based on the type II burst in the radio dynamic spectrum, we derived the altitude of the CME-driven shock to be 0.73 R$_{\odot}$ at 07:03 UT and 1.19 R$_{\odot}$ at 07:09 UT above the solar surface.
\end{itemize}

After being discovered in 1997  (\opencite{Thompson98}), EUV waves were initially thought to be fast-mode MHD waves or shock waves in the solar corona. However, the extremely low speeds (smaller than the coronal sound speed) in some events, and in particular, the discovery of stationary EUV wave front at magnetic QSLs (\opencite{Del99}), invoked \inlinecite{Chen02} to propose a two-wave scenario, \emph{i.e.} when a CME happens, there should exist two types of EUV waves, a fast-component EUV wave (which is a fast-mode wave or shock wave and corresponds to the coronal counterpart of the chromospheric Moreton wave) and a slow-component EUV wave (which is an apparent propagation produced by successive stretching of the closed magnetic field lines overlying the CME). The MHD numerical simulation performed by \inlinecite{Chen05} indeed showed that the slow-component EUV wave stops at magnetic QSLs, a natural result of the magnetic fieldline stretching model. However, \inlinecite{Del07} found that even the Moreton wave can also generate a stationary front in H$\alpha$. It implies that there should be two different mechanisms for the formation of stationary fronts. Although \inlinecite{Del07} already presented multiple stationary EUV wave fronts, they could not pin down the formation process owing to the low cadence of the EIT telescope. With the high cadence data observed by SDO/AIA, \inlinecite{Chandra16} revealed that when the fast-component EUV wave passes through a magnetic QSL, a stationary EUV wave front is generated. In order to explain the formation of this new kind of stationary wave front, \inlinecite{Chen16} proposed that before reaching a magnetic QSL, part of the fast-mode MHD wave in the corona is converted to a slow-mode MHD wave at the location where the Alfv\'en speed is equal to the sound speed. Since a slow-mode wave can only propagate along the magnetic field line until it decays at the footpoint of the field line, it is seen to be a stationary front as viewed from above. Since the slow-mode wave will finally be dissipated in the chromosphere, this model can also explain the stationary H$\alpha$ front discovered by \inlinecite{Del07}. Such a model was confirmed by later observations \citep{Zong17, Chandra18}. However, these observations displayed only a single stationary EUV wave front. In the real corona, a fast-mode EUV wave might pass through several magnetic QSLs, and therefore, it is expected to see the formation of a series of stationary EUV waves. In this paper, we showed that in the 2013 April 11 event, as the fast-mode EUV wave propagates outward to the east, two stationary EUV wave fronts are produced. The locations of the stationary fronts are exactly near magnetic QSLs.  The schematic representation of the observations and the formation of stationary fronts is shown in Figure \ref{cartoon}. 

Although \inlinecite{Del07} already showed the observation of multiple stationary EUV wave fronts, the low-cadence data did not allow them to pin down how these stationary fronts are formed. Besides, we proposed an alternative explanation for the formation of these stationary fronts. Since they are formed by the CME-driven shock wave, we tend to think that they are not related to the magnetic rearrangement, which happens much behind the shock wave.

\begin{figure} %%%%%%%%%%%%%%%%%% FIGURE 1
\begin{center}
\centerline{\hspace*{0.20\textwidth}
               \includegraphics[width=1.3\textwidth,clip=]{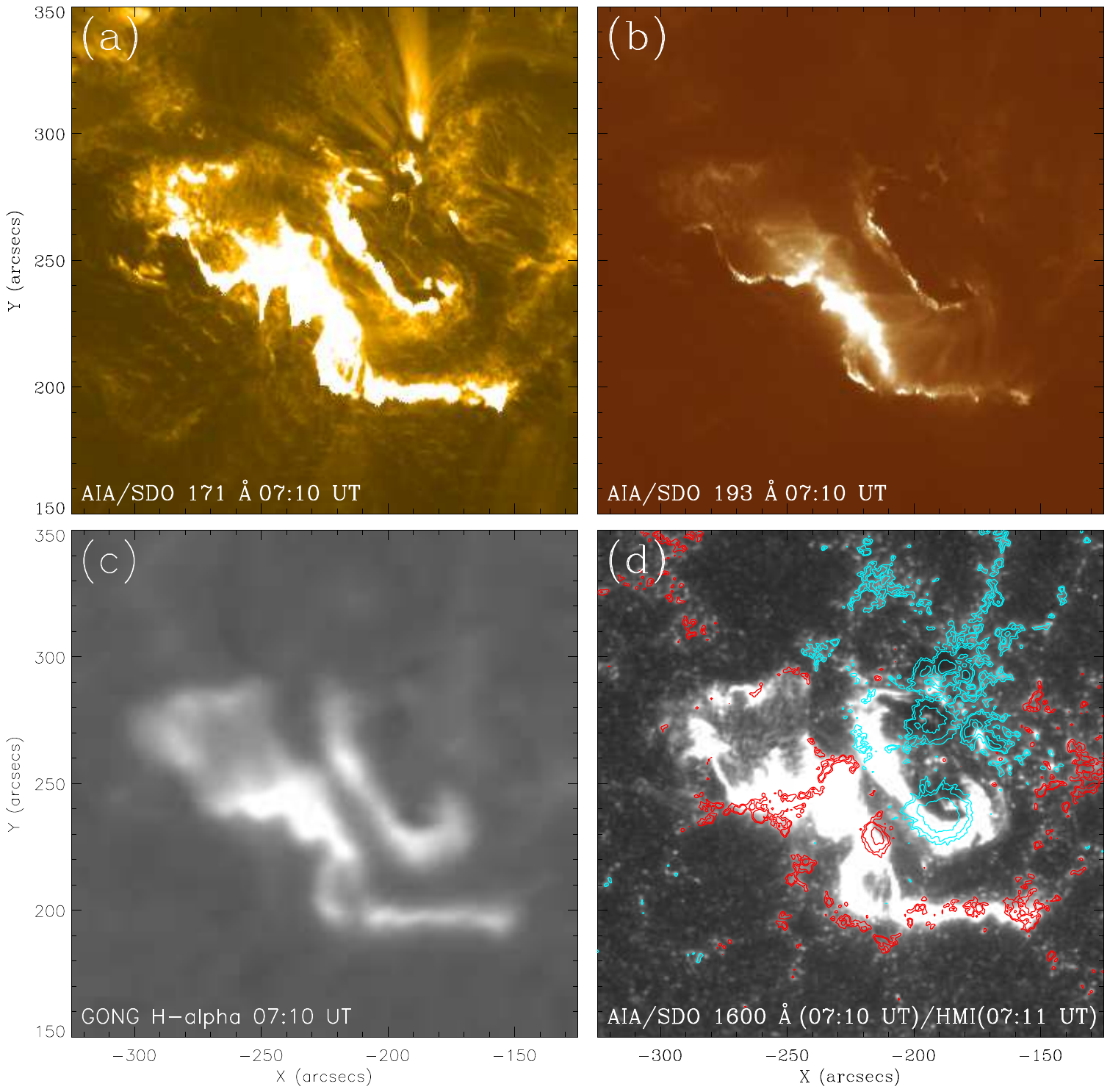}
              }
\vspace{-0.6\textwidth}    % Shift back to the panel bottom

\caption{Multiwavelength observations of the flare ribbons in AIA 171, 193,
1600 \AA{} and H$\alpha$ at 07:10 UT where the AIA 1600 \AA{} image is
overlaid by HMI contours (d). Red/cyan colors represent negative/positive
magnetic polarities respectively. The contour levels are ${\pm}200$, ${\pm}400$, ${\pm}800$, ${\pm}1600$ G.}
   \label{fla}
\end{center}
\end{figure}

\begin{figure} %%%%%%%%%%%%%%%%%% FIGURE 2
\begin{center}
%\vspace{-0.3\textwidth}    % Shift back to the panel bottom
\centerline{\hspace*{-0.01\textwidth}
               \includegraphics[width=1.0\textwidth,clip=]{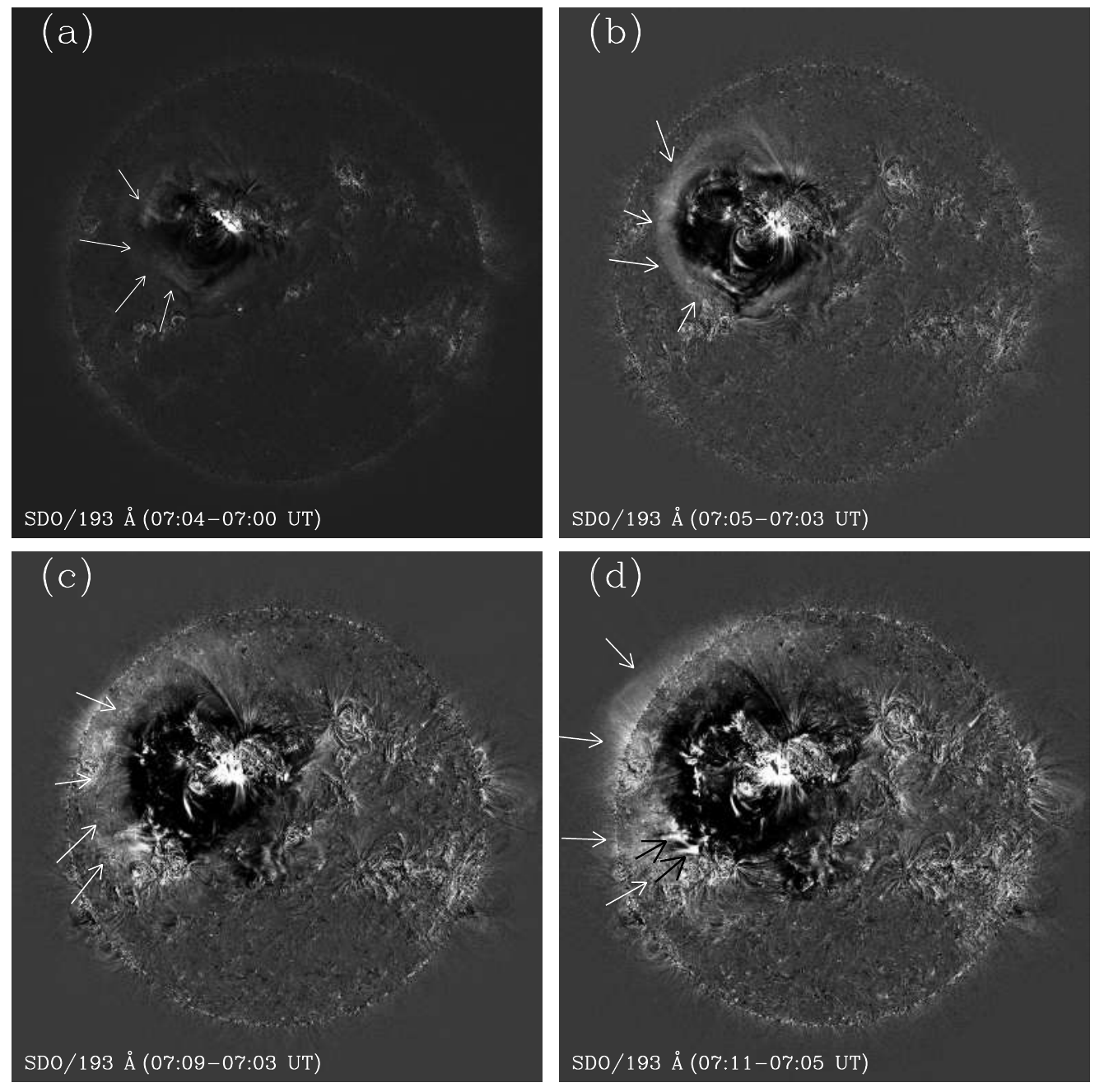}
              }
\vspace{-0.02\textwidth}    % Shift back to the panel bottom
\caption{Temporal and spatial evolution of the EUV wave in full disk running
difference images at AIA 193 \AA{}.
Arrows indicate the location of the propagating bright fronts. The white arrows
indicate the fast-mode MHD wave
and the black arrows indicate the slow-component EUV wave.}

   \label{wave}
\end{center}
   \end{figure}

It should also be noted that although the MHD simulations performed by \inlinecite{Chen02} displayed two types of EUV waves as a CME erupts, it does not mean that one can always see two different EUV waves along any direction. According to the magnetic fieldline stretching model, the slow-component EUV wave is generated by the stretching of the magnetic field lines straddling over the erupting flux rope. Therefore, the slow-component EUV wave appears only in the regions whose magnetic field lines straddle over the erupting flux rope. In contrast, the fast-component EUV wave, which is a CME-driven shock wave, would be more circular. As revealed in this paper, we can see only the fast-component EUV wave along the eastward direction (Slice 1) and two EUV waves are discernable along the southeast direction, which implies that the large-scale coronal magnetic field overlying the erupting flux rope is mainly along the southeast direction. Such an inference is consistent with the observational fact that the two ribbons of the associated flare, as indicated by Figure \ref{fla}, are separated in the southeast-northwest direction. Besides, according to the magnetic fieldline stretching model, the slow-component EUV wave would be $\approx$3 times slower than the fast-component EUV wave if the magnetic field lines are concentric semicircles. In this paper, the slow-component EUV wave is $\approx$4.3 times slower than the fast-component EUV wave, implying that the coronal magnetic field lines are more elongated in the solar radial direction \citep{Chen05}.

\begin{figure} %%%%%%%%%%%%%%%%%% FIGURE 3
\begin{center}
\vspace{-0.02\textwidth}    % Shift back to the panel bottom
\centerline{\hspace*{-0.05\textwidth}
               \includegraphics[width=1.0\textwidth,clip=]{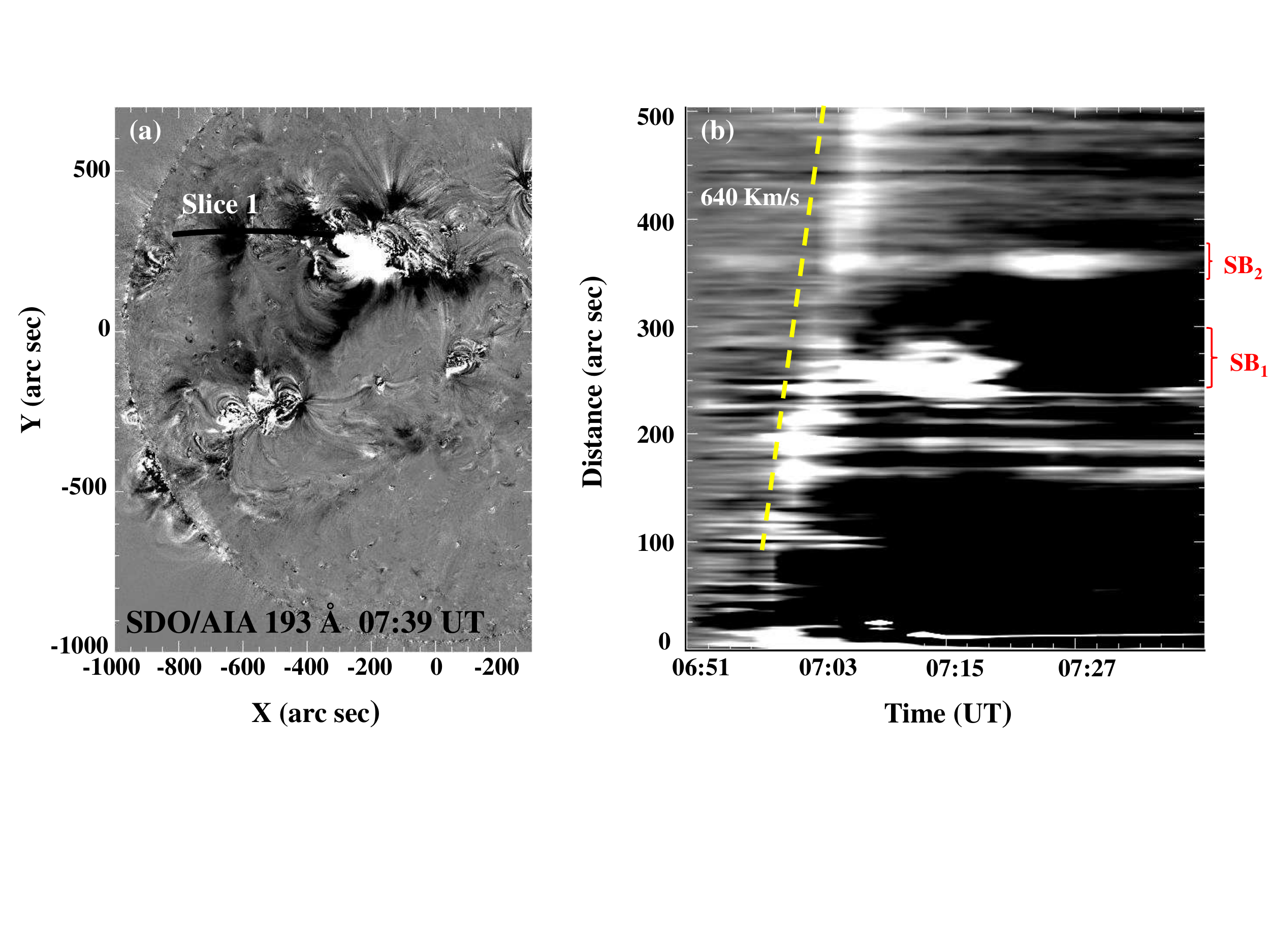}
              }
\vspace{-0.1\textwidth}    % Shift back to the panel bottom
\caption{(a): SDO/AIA 193 \AA\ difference image at 07:39 UT showing the direction of slice 1 (black line) used for the time--distance diagram in the right panel. (b): Time--distance diagram showing a fast-component EUV wave and two resulting stationary brightenings $SB_{1}$ and $SB_{2}$. }
%\vspace{-0.5\textwidth}

   \label{qsl}
\end{center}
   \end{figure}

\begin{figure} %%%%%%%%%%%%%%%%%% FIGURE 4
\begin{center}
\vspace{0.0\textwidth}    % Shift back to the panel bottom
\centerline{\hspace*{-0.03\textwidth}
               \includegraphics[width=1.5\textwidth,angle=0,clip=]{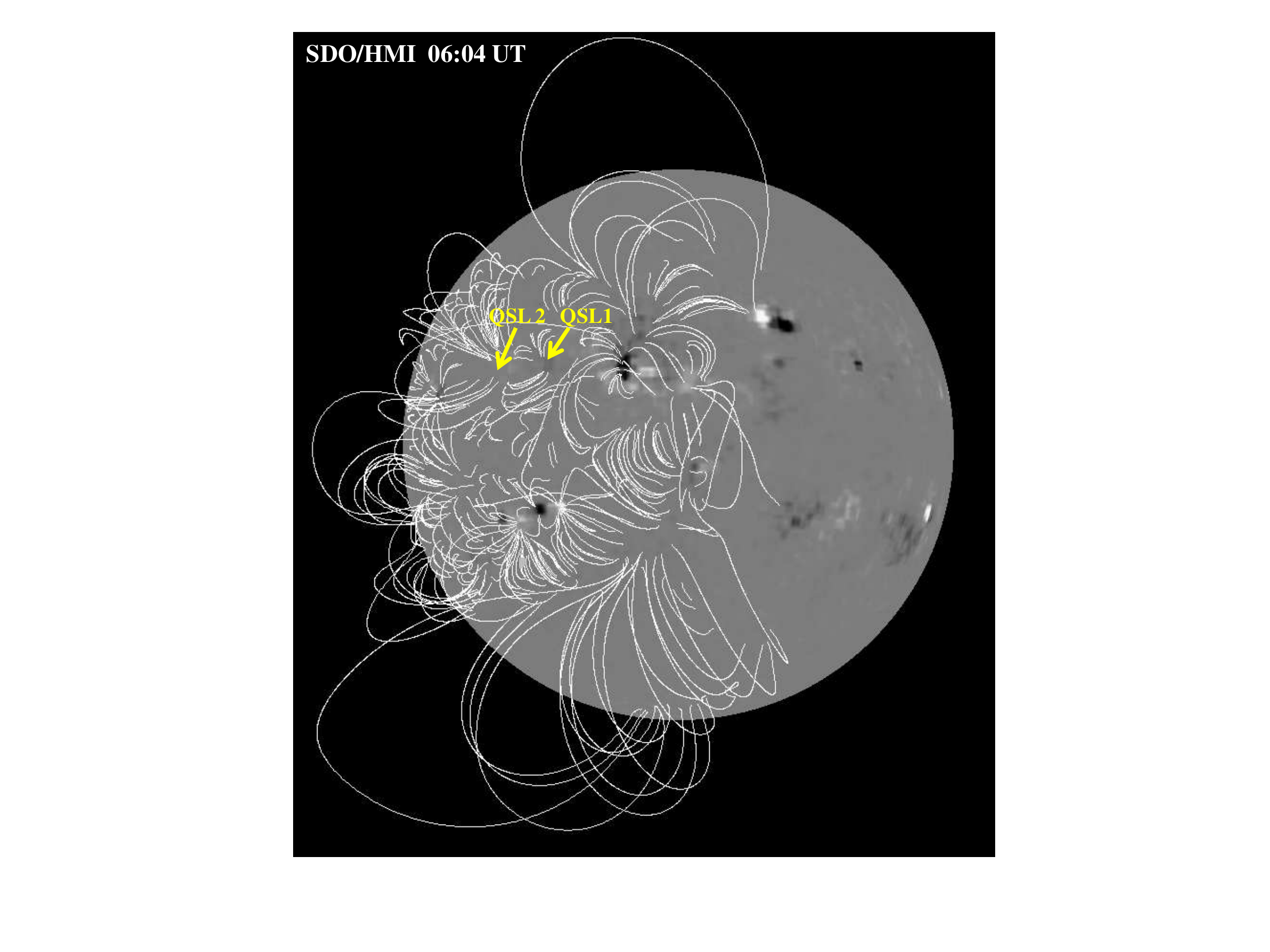}
              }
\vspace{-0.05\textwidth}    % Shift back to the panel bottom
\caption{Potential coronal magnetic field extrapolated from the HMI magnetic field at 06:04 UT. Yellow arrows indicate the locations of QSLs.}

   \label{pfss}
\end{center}
   \end{figure}

\begin{figure} %%%%%%%%%%%%%%%%%% FIGURE 5
\begin{center}
\vspace{-0.04\textwidth}    % Shift back to the panel bottom
\centerline{\hspace*{-0.05\textwidth}
               \includegraphics[width=1.0\textwidth,clip=]{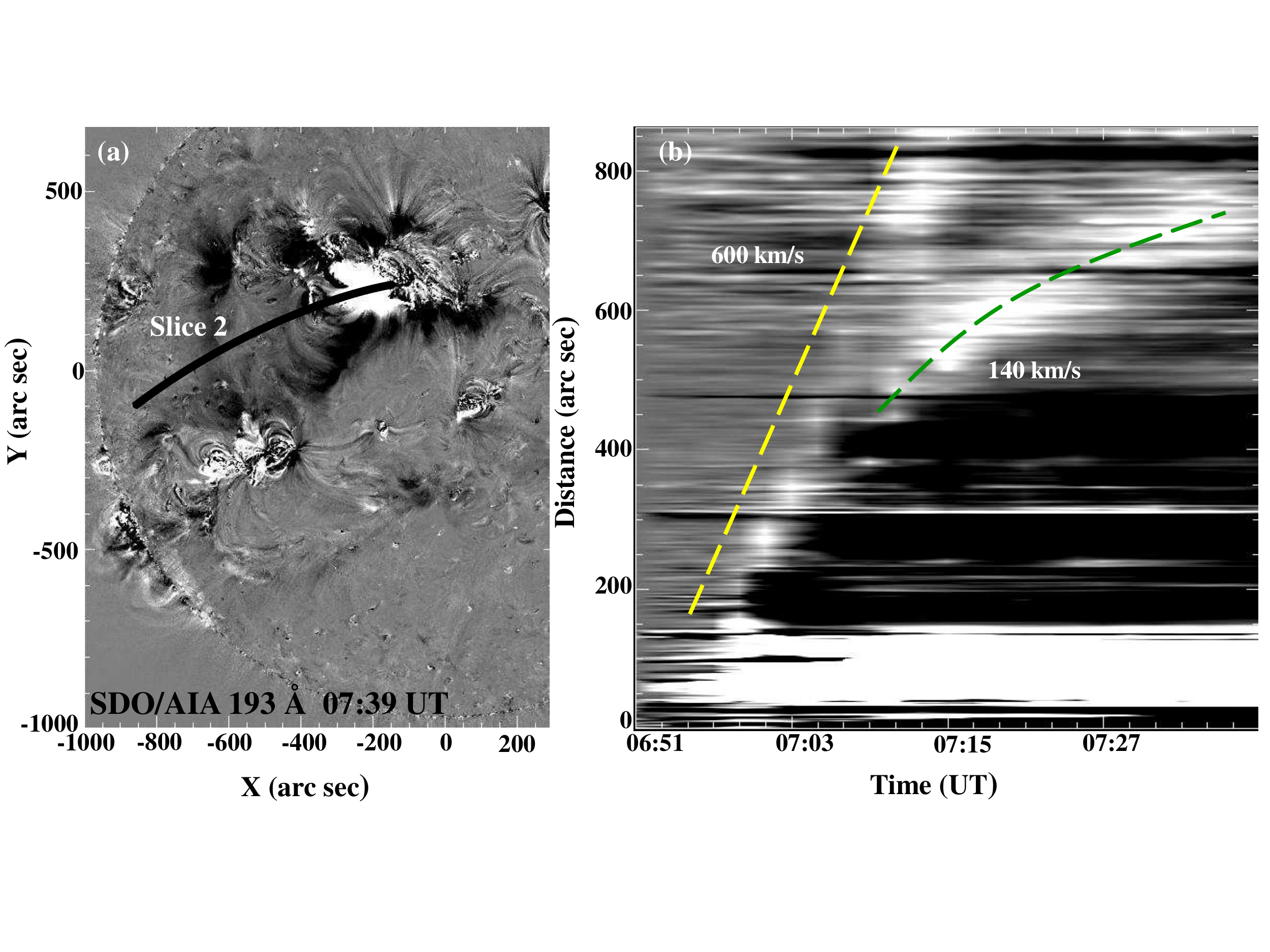}
              }
\vspace{-0.1\textwidth}    % Shift back to the panel bottom
\caption{(a): SDO/AIA 193 \AA\ difference image at 07:39 UT showing the direction of slice 2 (black line) used for the time--distance diagram in the right panel. (b): Time--distance diagram showing both the fast- and slow-component EUV waves.}
%\vspace{-0.5\textwidth}

  \label{twowave}
\end{center}
   \end{figure}

\begin{figure} %%%%%%%%%%%%%%%%%% FIGURE 6
\begin{center}
\vspace{0.0\textwidth}    % Shift back to the panel bottom
\centerline{\hspace*{0.0\textwidth}
               \includegraphics[width=1.0\textwidth,clip=]{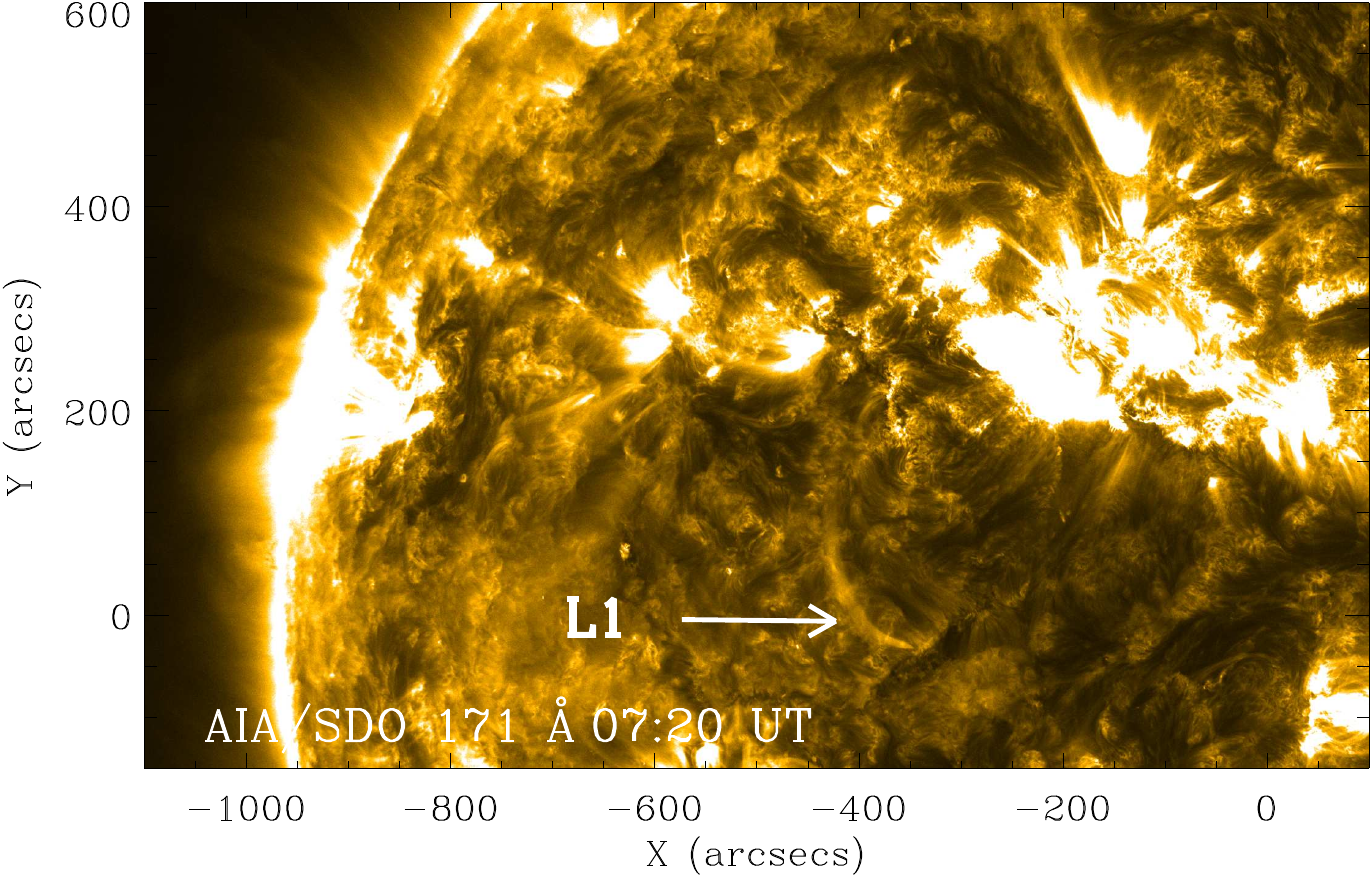}
              }
%\vspace{-0.05\textwidth}    % Shift back to the panel bottom
\caption{AIA/SDO 171~\AA{} image showing the oscillating coronal loop $L_{1}$
near the flare site. }

   \label{loop}
\end{center}
   \end{figure}

\begin{figure} %%%%%%%%%%%%%%%%%% FIGURE 7
\begin{center}
\vspace{0.02\textwidth}    % Shift back to the panel bottom
$\color{black} \put(-145,-300){\Large \bf (a)} $
$\color{black} \put(-20,-300){\Large \bf (b)} $

\centerline{\hspace*{0.0\textwidth}
               \includegraphics[width=1.0\textwidth,clip=]{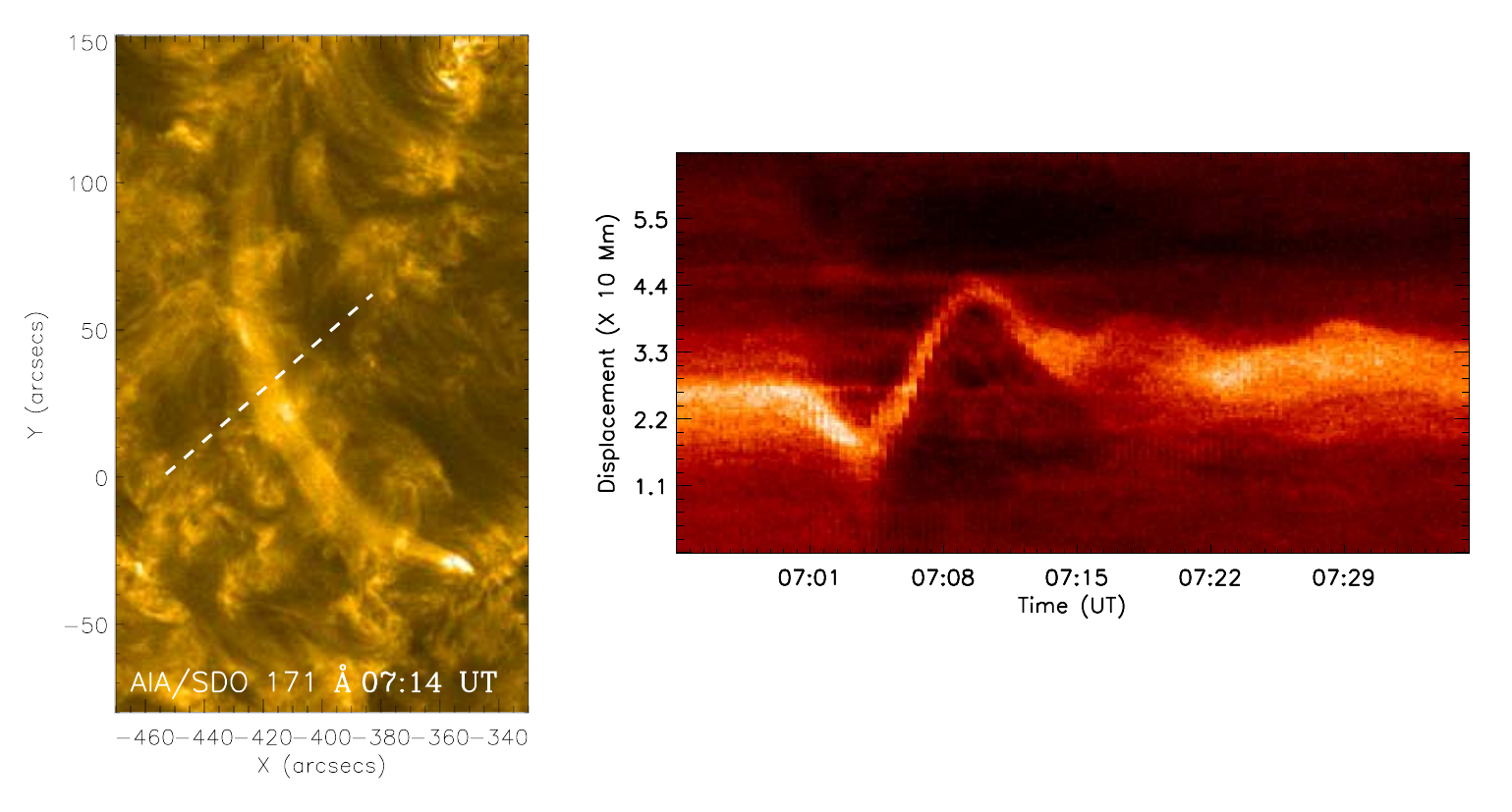}
              }
%\vspace{-0.09\textwidth} 
   % Shift back to the panel bottom
\caption{(a): Oscillating loop in AIA/SDO 171~\AA{}. The location of the artificial slit used for the timeslice analysis is shown by the white dashed line.
(b): The time--distance plot of the oscillations of the loop along the slit
shown in the left panel.}
%\vspace{-0.09\textwidth}
   \label{slice}
\end{center}
   \end{figure}

\begin{figure} %%%%%%%%%%%%%%%%%% FIGURE 8
\begin{center}
%\vspace{2.0\textwidth}    % Shift back to the panel bottom
\centerline{\hspace*{0.09\textwidth}
               \includegraphics[width=1.2\textwidth,angle=180,clip=]{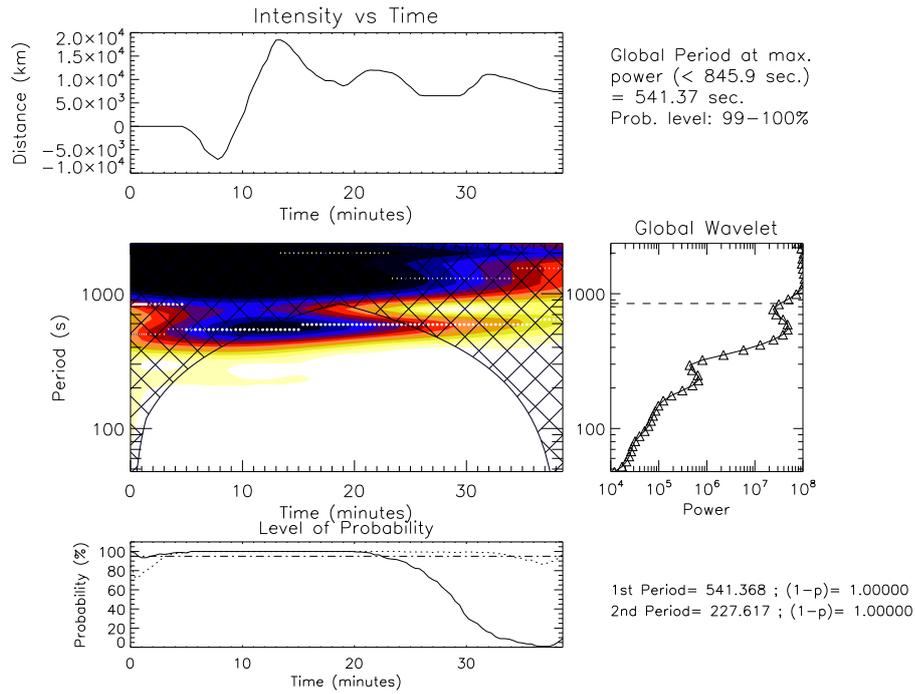}
              }
\vspace{-0.01\textwidth}    % Shift back to the panel bottom
\caption{The wavelet result for the loop oscillations in AIA 171 \AA{}. The
top panel shows the intensity variation with time in AIA 171 \AA{}.
The wavelet power spectrum is shown in the middle panel, and the probability is given in the bottom panel.}

   \label{wavelet}
\end{center}
   \end{figure}

\begin{figure} %%%%%%%%%%%%%%%%%% FIGURE 9
\begin{center}
\vspace{-0.0\textwidth}    % Shift back to the panel bottom
\centerline{\hspace*{-0.01\textwidth}
               \includegraphics[width=1.0\textwidth,clip=]{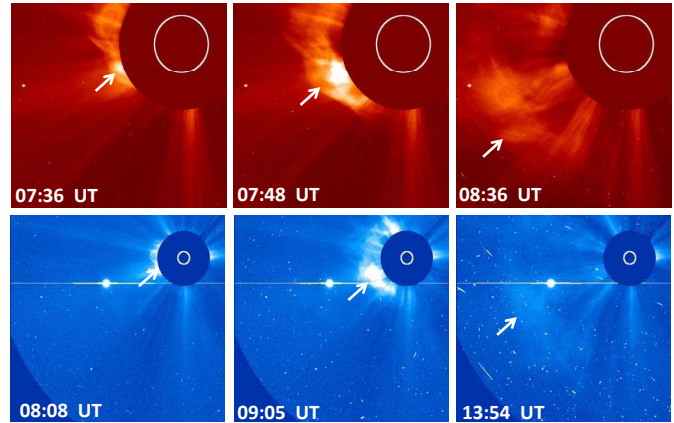}
              }
\vspace{-0.15\textwidth}    % Shift back to the panel bottom

\caption{Evolution of the associated CME observed by the LASCO C2 (top) and C3 (bottom) coronagraphs.}

   \label{cme}
\end{center}
   \end{figure}

\begin{figure} %%%%%%%%%%%%%%%%%% FIGURE 10
\begin{center}

\vspace{-0.0\textwidth}    % Shift back to the panel bottom
\centerline{\hspace*{-0.05\textwidth}
               \includegraphics[width=0.8\textwidth,clip=]{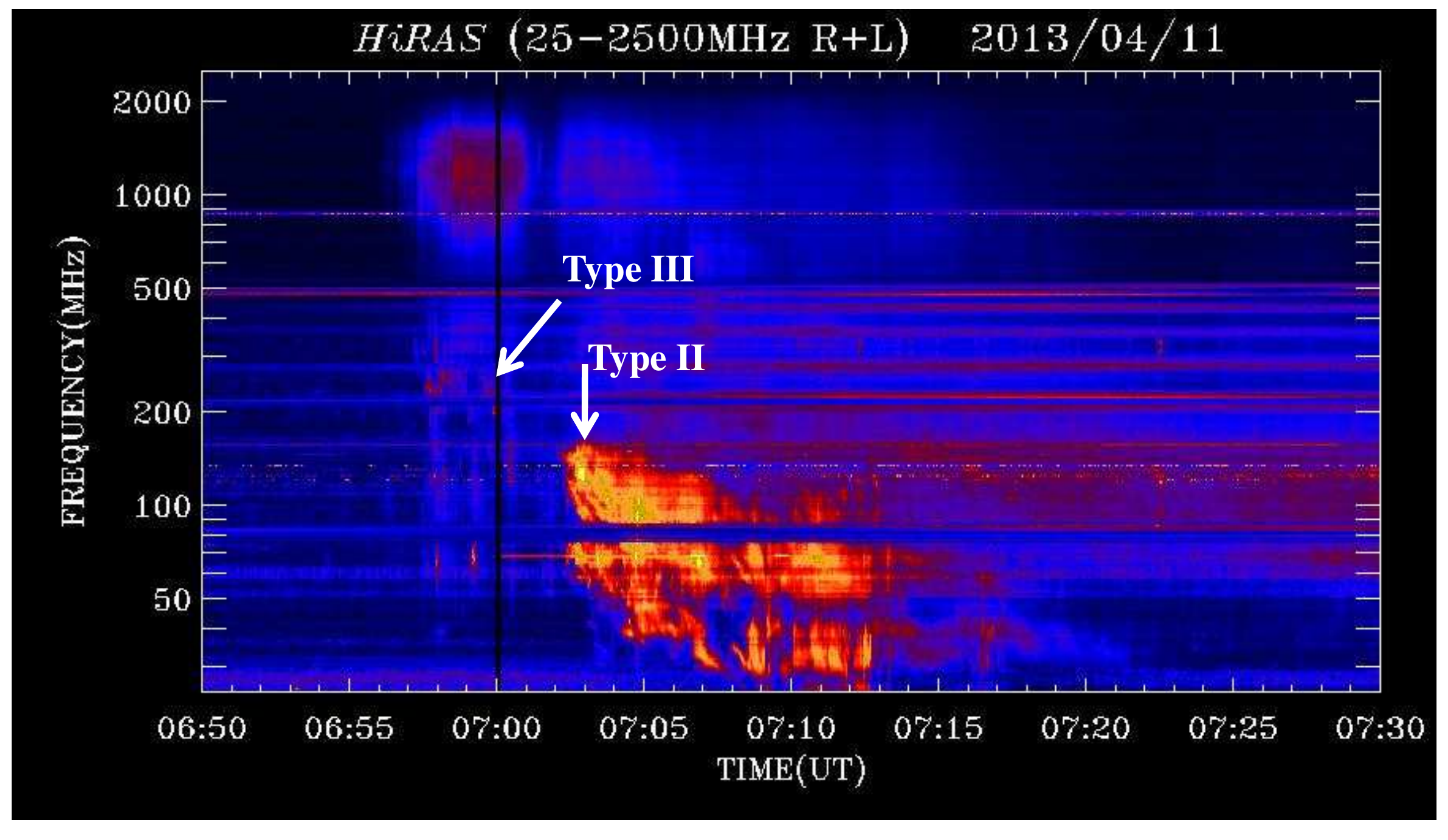}
              }
\vspace{0.0\textwidth}    % Shift back to the panel bottom
\caption{The dynamic spectrum observed by the Hiraiso Radio Spectrograph (HiRAS) in 25--2500 MHz on 2013 April 11 showing type II and type III radio bursts during the flare/CME event.}

   \label{type}
\end{center}
   \end{figure}

\begin{figure} %%%%%%%%%%%%%%%%%% FIGURE 11
\begin{center}
\vspace{0.0\textwidth}    % Shift back to the panel bottom
\centerline{\hspace*{-0.03\textwidth}
               \includegraphics[width=1.0\textwidth,clip=]{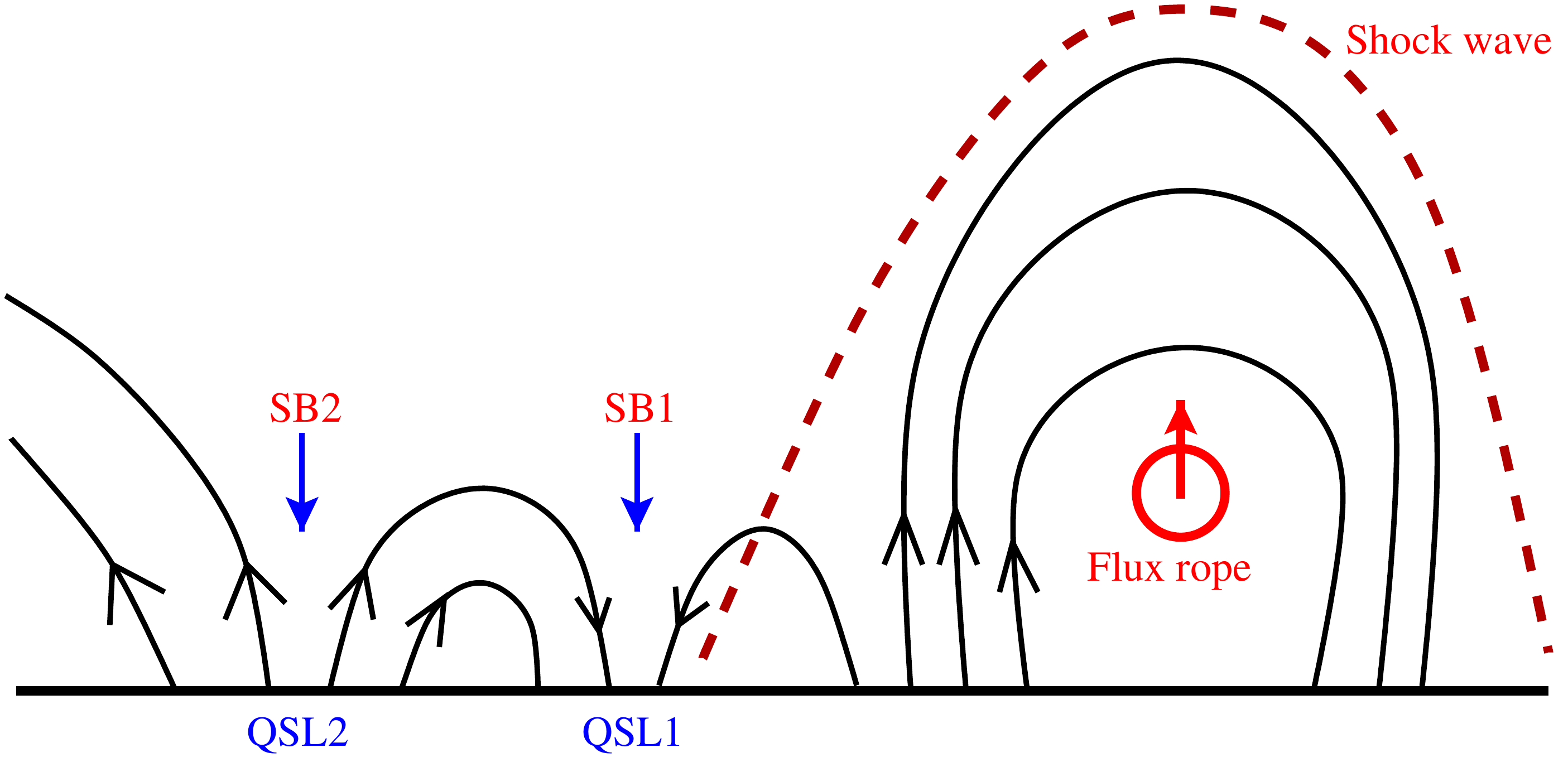}
              }
\vspace{0.0\textwidth}    % Shift back to the panel bottom
\caption{Cartoon showing the propagation of the fast mode EUV wave/shock wave (red dashed arc). The black solid lines represent the initial magnetic configuration. The fast-mode EUV wave encounters two QSLs, producing two stationary brightenings.} \label{cartoon}
\end{center} 
   \end{figure}

According to the magnetic fieldline stretching model, the fast-component EUV wave and the type II burst source region represent different parts of the same CME piston-driven shock wave. Therefore, the locations of the fast-component EUV wave and the type II radio bursts can provide the shape information of the CME-driven shock wave when it can not be directly imaged as in \inlinecite{Ma11}. With the Newkirk 2-fold density model, the radio burst is at the altitude of 0.73 R$_{\odot}$ above the solar surface at 07:03 UT. At this moment, the fast-component EUV wave is 350$^{\prime \prime}$ (\emph{i.e.} 0.37 R$_{\odot}$) away from the flare site according to Figure \ref{qsl}, implying that the shock wave is relatively elongated in the solar radial direction.
It is noted in passing that we found that the fast-component EUV wave appears 5 minutes before the onset of the type II radio burst. This result is of significance in clarifying the debate about whether the commencement of type II radio burst is indicative of the formation of a shock wave. Our observation implies that only when the shock is strong enough, \emph{e.g.} a supercritical shock (\opencite{Benz88}), can a type II radio burst be excited (\opencite{Su16}).

Similar to the study of \inlinecite{Ballai05} and \inlinecite{Ballai07}, we observed loop oscillations as the fast-component EUV wave impinges the coronal loop. To estimate the total energy of the EUV wave, we adopted the computation method proposed by the above authors. Different from \inlinecite{Ballai05} who used typical coronal physical parameters in their calculation, we derived the input physical parameters using the SDO/AIA data. For the EUV wave event studied in this paper, the total energy is estimated to be $2.7\times10^{20}$~J.

\begin{acks}
We would like to thank the referee for the useful comments and suggestions that helped us to improve the manuscript. We also acknowledge the use of SDO and GONG data. PFC was supported by the Chinese grants NSFC 11533005, U1731241 and Jiangsu 333 Project (No. BRA2017359). AF acknowledges the support from the ISRO/RESPOND project. The work of IZh and RC was supported by the Bulgarian Science Fund under Indo--Bulgarian bilateral project DNTS/INDIA 01/7. 

\vspace{3mm}

\noindent {\bf Disclosure of Potential Conflicts of Interest}  The authors declare that they have no conflicts of interest.

\end{acks}

\mbox{}~\\
\bibliographystyle{spr-mp-sola}
\bibliography{reference}
\IfFileExists{\jobname.bbl}{} {\typeout{}
\typeout{***************************************************************}
\typeout{***************************************************************}
\typeout{** Please run "bibtex \jobname" to obtain the bibliography}
\typeout{** and re-run "latex \jobname" twice to fix references}
\typeout{***************************************************************}
\typeout{***************************************************************}
\typeout{}}

%%%%%%%%%%%%%%%%%%%%%%%%%%%%%%%%%%%%%%%%%%%%%%%%%%%%%%%%%%%%%%%%%%%%%%%%%%%%%%%%%%%%%%%%%%%%%%%%%%%%%%%%%%%%%%

%%%%%%%%%%%%%%%%%%%%%%%%%%%%%%%%%%%%%%%%%%%%%%%%%%%%%%%%%%%%%%%%%%%%%%%%%%%%%%%%%%%%%%%%%%%%%%%%%%%%%%%

%%%%%%%%%%%%%%%%%%%%%%%%%%%%%%%%%%%%%%%%%%%%%%%%%%%%%%%%%%%%%%%%%%%%%%%%%%%%%%%%%%%%%%%%%%%%%%%%%%%%%%%%

%%%%%%%%%%%%%%%%%%%%%%%%%%%%%%%%%%%%%%%%%%%%%%%%%%%%%%%%%%%%%%%%%%%%%%%%%%%%%%%%%%%%%%%%%%%%%%%%%%%%%

\clearpage

%%%%%%%%%%%%%%%%%%%%%%%%%%%%%%%%%%%%%%%%%%%%%%%%%%%%%%%%%%%%%%%%%%%%%%%%%%%%%%%%%%%%%%%%%%%%%%%%%%%%%%%%

%%%%%%%%%%%%%%%%%%%%%%%%%%%%%%%%%%%%%%%%%%%%%%%%%%%%%%%%%%%%%%%%%%%%%%%%%%%%%%%%%%%%%%%%%%%%%%%%%%%%%
 \clearpage

%%%%%%%%%%%%%%%%%%%%%%%%%%%%%%%%%%%%%%%%%%%%%%%%%%%%%%%%%%%%%%%%%%%%%%%%%%%%%%%%%%%%%%%%%%%%%%%%%%%%%%

%%%%%%%%%%%%%%%%%%%%%%%%%%%%%%%%%%%%%%%%%%%%%%%%%%%%%%%%%%%%%%%%%%%%%%%%%%%%%%%%%%%%%%%%%%%%%%%%%%%%%

\vspace{-5.0\textwidth}
%%%%%%%%%%%%%%%%%%%%%%%%%%%%%%%%%%%%%%%%%%%%%%%%%%%%%%%%%%%%%%%%%%%%%%%%%%%%%%%%%%%%%%%%%%%%%%%%%%%%%%%

%%%%%%%%%%%%%%%%%%%%%%%%%%%%%%%%%%%%%%%%%%%%%%%%%%%%%%%%%%%%%%%%%%%%%%%%%%%%%%%%%%%%%%%%%%%%%%%%%%%%%
 \clearpage

%%%%%%%%%%%%%%%%%%%%%%%%%%%%%%%%%%%%%%%%%%%%%%%%%%%%%%%%%%%%%%%%%%%%%%%%%%%%%%%%%%%%%%%%%%%%%%%%%%%%%%

%%%%%%%%%%%%%%%%%%%%%%%%%%%%%%%%%%%%%%%%%%%%%%%%%%%%%%%%%%%%%%%%%%%%%%%%%%%%%%%%%%%%%%%%%%%%%%%%%%%%%
 \clearpage

\vspace{-5.0\textwidth}
%%%%%%%%%%%%%%%%%%%%%%%%%%%%%%%%%%%%%%%%%%%%%%%%%%%%%%%%%%%%%%%%%%%%%%%%%%%%%%%%%%%%%%%%%%%%%%%%%%%%%%%

%%%%%%%%%%%%%%%%%%%%%%%%%%%%%%%%%%%%%%%%%%%%%%%%%%%%%%%%%%%%%%%%%%%%%%%%%%%%%%%%%%%%%%%%%%%%%%%%%%%%%%
\vspace{-0.6\textwidth}

%%%%%%%%%%%%%%%%%%%%%%%%%%%%%%%%%%%%%%%%%%%%%%%%%%%%%%%%%%%%%%%%%%%%%%%%%
\clearpage
%%%%%%%%%%%%%%%%%%%%%%%%%%%%%%%%%%%%%%%%%%%%%%%%%%%%%%%%%%%%%%%%%%%%%%%%%

%%%%%%%%%%%%%%%%%%%%%%%%%%%%%%%%%%%%%%%%%%%%%%%%%%%%%%%%%%%%%%%%%%%%%%%%%%%%%%%%%%%%%%%%%%%%%%%%%%%%
\vspace{-8.0\textwidth}
%%%%%%%%%%%%%%%%%%%%%%%%%%%%%%%%%%%%%%%%%%%%%%%%%%%%%%%%%%%%%%%%%%%%%%%%%%%%%%%%%%%%%%%%%%%%%%%%%%%%%%%%

%%%%%%%%%%%%%%%%%%%%%%%%%%%%%%%%%%%%%%%%%%%%%%%%%%%%%%%%%%%%%%%%%%%%%%%%%%%%%%%%%%%%%%%%%%%%%%%%%%%%%

\clearpage
%%%%%%%%%%%%%%%%%%%%%%%%%%%%%%%%%%%%%%%%%%%%%%%%%%%%%%%%%%%%%%%%%%%%%%%%%%%%%%%%%%%%%%%%%%%%%%%%%%%%%%%

%%%%%%%%%%%%%%%%%%%%%%%%%%%%%%%%%%%%%%%%%%%%%%%%%%%%%%%%%%%%%%%%%%%%%%%%%%%%%%%%%%%%%%%%%%%%%%%%%%%%%

\clearpage

%%%%%%%%%%%%%%%%%%%%%%%%%%%%%%%%%%%%%%%%%%%%%%%%%%%%%%%%%%%%%%%%%%%%%%%%%%%%%%%%%%%%%%%%%%%%%%%%%%%%%%%%
%%%%%%%%%%%%%%%%%%%%%%%%%%%%%%%%%%%%%%%%%%%%%%%%%%%%%%%%%%%%%%%%%%%%%%%%%%%%%%%%%%%%%%%%%%%%%%%%%%%%%

\vspace{-0.2\textwidth}
%%%%%%%%%%%%%%%%%%%%%%%%%%%%%%%%%%%%%%%%%%%%%%%%%%%%%%%%%%%%%%%%%%%%%%%%%%%%%%%%%%%%%%%%%%%%%%%%%%%%%%%%

%%%%%%%%%%%%%%%%%%%%%%%%%%%%%%%%%%%%%%%%%%%%%%%%%%%%%%%%%%%%%%%%%%%%%%%%%%%%%%%%%%%%%%%%%%%%%%%%%%%%%
%\clearpage

%%%%%%%%%%%%%%%%%%%%%%%%%%%%%%%%%%%%%%%%%%%%%%%%%%%%%%%%%%%%%%%%%%%%%%%%%%%%%%%%%%%%%%%%%%%%%%%%%%%%%

%
\clearpage

%%%%%%%%%%%%%%%%%%%%%%%%%%%%%%%%%%%%%%%%%%%%%%%%%%%%%%%%%%%%%%%%%%%%%%%%%%%%%%%%%%%%%%%%%%%%%%%%%%%%%%%%%%%%%%

\end{article}
\end{document}